%% file: 0.main.tex
\documentclass[npa,12pt]{elsarticle}  

\usepackage{graphicx}
\usepackage{lineno}
\usepackage{hyperref}
\usepackage[utf8]{inputenc}
\usepackage{multirow}
\usepackage{booktabs}
\usepackage{rotating}
\usepackage{orcidlink}

\newcommand{\CAMOO}{$^{48\mathrm{depl}}$Ca$^{100}$MoO$_{4}${}}

\def\a{$\alpha${}}
\def\b{$\beta${}}
\newcommand{\g}{$\gamma${}}

\newcommand{\sm}{$\sim${}}
\def\lt{$<$}

\def\znbb{$0\nu\beta\beta$}
\def\tnbb{$2\nu\beta\beta$}

\newcommand\geniso[2]{\ensuremath{\rm ^{#2}#1}}

\newcommand\newiso[3][]{
\ifx&#1&
  \expandafter\newcommand\csname #2\endcsname[1][#3]{\geniso{#2}{##1}}
\else
  \expandafter\newcommand\csname #1\endcsname[1][#3]{\geniso{#2}{##1}}
\fi
}

\newiso{Ca}{48}
\newiso{K}{40}
\newiso{Th}{232}
\newiso{Pa}{231}
\newiso{U}{238}
\newiso{Rn}{222}
\newiso{Ra}{226}
\newiso{Pb}{208}
\newiso{Tl}{208}
\newiso{Bi}{214}
\newiso{Ac}{226}
\newiso{Po}{210}
\newiso{Mo}{100}
\newiso{Co}{60}

\begin{document}

\begin{frontmatter} 
\input{amore_author_orcid}
\title{Background study of the AMoRE-pilot experiment}

\begin{abstract}
We report a study on the background of the Advanced Molybdenum-Based Rare process Experiment (AMoRE), a search for neutrinoless double beta decay (\znbb) of $^{100}$Mo. 
The pilot stage of the experiment was conducted using $\sim$1.9 kg of \CAMOO~ crystals at the Yangyang Underground Laboratory, South Korea, from 2015 to 2018. We compared the measured $\beta/\gamma$ energy spectra in three experimental configurations with the results of Monte Carlo simulations and identified the background sources in each configuration. We replaced several detector components and enhanced the neutron shielding to lower the background level between configurations.
A limit on the half-life of $0\nu\beta\beta$ decay of $^{100}$Mo was found at $T_{1/2}^{0\nu} \ge 3.0\times 10^{23}$ years at 90\% confidence level, based on the measured background and its modeling. Further reduction of the background rate in the AMoRE-I and AMoRE-II are discussed. 
\end{abstract}


\end{frontmatter}


\input{1.intro-v2}

\input{2.experiment}
\input{3.dataAnalysis}
\input{4.bkgSrcs}
\input{5.simulation}

\input{6-1.result}
\input{6-2.discussion}
\input{7.summary}

\section*{Acknowledgments}

This research was funded by the Institute for Basic Science (Korea) under project codes IBS-R016-D1 and IBS-R016-A2. It is also supported by the Ministry of Science and Higher Education of the Russian Federation (N121031700314-5), the MEPhI Program Priority 2030, the National Research Foundation of Korea (NRF-2021R1I1A3041453, NRF-2021R1A2C1013761 NRF-2018K1A3A1A13087769), and the National Research Facilities \& Equipment Center (NFEC) of Korea (No. 2019R1A6C1010027).

These acknowledgements are not to be interpreted as an endorsement of any statement made by any of our institutes, funding agencies, governments, or their representatives.

\bibliographystyle{spphys}
\bibliography{reference}

\end{document}

%% file: amore_author_orcid.tex
\author{A.~Agrawal\,\orcidlink{0000-0001-7740-5637}}
\author{V.V.~Alenkov\,\orcidlink{0009-0008-8839-0010}}
\author{P.~Aryal\,\orcidlink{0000-0003-4955-6838}}
\author{J.~Beyer\,\orcidlink{0000-0001-9343-0728}}
\author{B.~Bhandari\,\orcidlink{0009-0009-7710-6202}}
\author{R.S.~Boiko\,\orcidlink{0000-0001-7017-8793}}
\author{K.~Boonin\,\orcidlink{0000-0003-4757-7926}}
\author{O.~Buzanov\,\orcidlink{0000-0002-7532-5710}}
\author{C.R.~Byeon\,\orcidlink{0009-0002-6567-5925}}
\author{N.~Chanthima\,\orcidlink{0009-0003-7774-8367}}
% ...


%% file: 1.intro-v2.tex
\section{Introduction}
\label{sec:intro}

Neutrinos are found to be massive, as shown by oscillation experiments with solar neutrinos and atmospheric neutrinos. The three mixing angles and two mass differences are all measured with the various oscillation experiments. Though the absolute masses of neutrinos are not measured yet, we know the neutrinos have very small masses compared with the lightest particle, the electron. 

The KATRIN group reported the result of a tritium beta decay experiment which showed that the mass of electron neutrinos is less than $0.8\ {\rm eV}/c^{2}$ \cite{KATRIN2022}. 
The neutrino mass can also be estimated using astrophysical observations: measuring the fluctuations in the temperature of the cosmic microwave background (CMB), mapping out the Universe structure by extensive galaxy surveys, and using observational Hubble parameter data \cite{Giusarma:2013pmn,Moresco:2012by,Parkinson:2012vd,Abazajian:2011dt}.
Regardless of the mixing angles, the total neutrino mass affects both the CMB radiation and the matter power spectra in the Universe, as well as effective mass values \cite{Beringer:1900zz}.
Therefore, their absolute mass scale 
plays a crucial role in the construction of the Universe, and conversely, neutrino properties, specifically their total mass and effective mass values, can be constrained by utilizing cosmic observational data.

The extremely small neutrino mass has not yet been understood, but the well-known seesaw mechanism suggests that the small masses of active neutrinos may originate from the heavy right-handed sterile neutrinos \cite{Mohapatra:1979ia}. Generally, the seesaw mechanism requires that the neutrinos are Majorana particles, neutrinos are the same as their anti-neutrinos and violate the lepton number conservation. 
Another important question is that matter dominates over anti-matter in the Universe. The cause of this asymmetry has not been understood. A possible solution to this puzzle that has been suggested in the neutrino sector is called leptogenesis \cite{Fukugita:1986hr,Deppisch:2018}. 
Even if charge-parity (CP) is observed to be violated in neutrino oscillation experiments, the theory for a leptogenesis-induced matter-antimatter asymmetry depends on whether neutrinos are Majorana particles or not.  

To confirm the Majorana nature of neutrinos, observing neutrinoless double beta (\znbb) decay has been suggested, and it is considered the most promising method for that purpose \cite{Giunti:2007ry,Mohapatra:2005wg}. Assuming that \znbb~decay occurs via the exchange of a light Majorana $\nu$, the effective Majorana mass is derived as 
\begin{equation}
m_{\beta \beta} = \left| \sum_{i=1}^{3} U_{ei}^{2} m_{i} \right|,    
\end{equation}
where $U_{ei}$ is the Pontecorvo-Maki-Nakagawa-Sakata (PMNS) matrix components for mixing between $i{\rm th}$ mass and electron flavor eigenstates \cite{Doi:1985dx}. Further, this effective Majorana mass can be  obtained experimentally from the measured half-life of \znbb decay.

 In particular, the effective Majorana mass depends on the ordering of the neutrino masses as normal or inverted, including the possibility of having a vanishing effective Majorana mass for the normal ordering, potentially resulting from destructive interferences driven by the Majorana phases.
In recent neutrino oscillation experiments at long baselines, normal mass ordering is preferred at the 3$\sigma$ level. The neutrino mass ordering will be decided with higher sensitivity by the upcoming reactor and long-baseline neutrino experiments\cite{An:2015jdp, DUNE:2021mtg, Kudenko:2020snj}. 
In short, the discovery of \znbb~decay will confirm lepton-number violation and help to determine the absolute neutrino mass scale and nature of neutrinos, which will be critical to understanding matter-antimatter asymmetry and cosmological observations of the Universe. 

The AMoRE experiment searches for the \znbb~decay of 100 kg of \Mo~nuclei using molybdate-based crystals operating at milli-Kelvin (mK) temperature. The $Q$-value of $^{100}$Mo double-beta decay has been reported to be $3034.40 \pm 0.17$ keV \cite{Rahaman:2007ng}.
The experiment aims to achieve zero background in the region of interest (ROI, $3034 \pm 7$ keV in AMoRE-pilot), so that the background counts inferred from the side-band data should be less than an order of one for the duration of the experiment of five years. In this case, the experimental sensitivity for the limit on the \znbb~decay half-life ($T_{1/2}^{0\nu}$) increases with the experiment exposure linearly as in the following equation.
\begin{equation}
\label{eq:livetimezbg}
    T_{1/2}^{0\nu} \sim \frac{N_{A} \cdot a \cdot \varepsilon}{A} \cdot \frac{M\cdot t}{ n_{CL}}, 
\end{equation}
where $N_{A}$ is the Avogadro’s number,
$a$ is the concentration of the isotope of interest,
$\varepsilon$ is the detection efficiency,
$A$ is the atomic mass of the \znbb~candidate nuclei, 
$M$ is the total detector mass, 
$t$ is the exposure time, 
and $n_{CL}$ is the number of events that can be excluded at a given confidence level (2.44 for 90\% C.L. according to \cite{Feldman1998}).
If the background event rate is non-zero, the sensitivity can be estimated as:
\begin{equation}
\label{eq:livetimebg}
    T_{1/2}^{0\nu} \sim \frac{N_{A} \cdot a \cdot \varepsilon}{A} \cdot \sqrt{\frac{M\cdot t}{b\cdot \Delta E_{\rm ROI}}},
\end{equation}
where $b$ is the rate of background events in the unit energy at ROI for the unit mass$\cdot$time exposure of detector material (counts/keV/kg/year, ckky), and $\Delta E_{\rm ROI}$ is the energy range for the signal, which is related to the detector energy resolution.

Based on the results obtained in our previous study with the 111~kg$\cdot$day exposure using \CAMOO\ (CMO) crystals, we have reported a background rate of 0.55 ckky in the energy range of 2850-3150 keV, and the corresponding $0\nu\beta\beta$ decay half-life limit of $T_{1/2}^{0\nu} > 9.5\times 10^{22}$ years at 90\% confidence level~\cite{Alenkov:2019jis}. Here, we report the updated and final result of the analysis on all available data from the AMoRE-pilot stage. The experimental apparatus is described in Section~\ref{sec:experiment}. Event selections and data analysis are explained in Section~\ref{sec:data}. In Sections~\ref{sec:bkgsrcs} and \ref{sec:sim}, we analyze the background energy spectrum in light of Monte Carlo simulations interpreted using radioassay of various detector components. Finally, a new limit on $T_{1/2}^{0\nu}$ is derived thanks to a better understanding of the background sources in the AMoRE.

%% file: 2.experiment.tex
\section{AMoRE-pilot Experiment}
\label{sec:experiment}

The AMoRE is following a three-stage plan consisting of AMoRE-pilot, AMoRE-I, and AMoRE-II. In the pilot stage, which is the topic of this work, the experiment was conducted at the Yangyang Underground Laboratory (Y2L) located at approximately 700~m vertical depth from the ground surface~\cite{Yoon:2021tkv}. We took data from 2015 to 2018 with about 1.9~kg of CMO crystals.  
Since the AMoRE-pilot experimental apparatus was described in detail in our previous report~\cite{Alenkov:2019jis}, here we summarize the key features and explain the changes between the experimental configurations\footnote{Called "runs" in the previous report.} after that.

\subsection{Detectors and shields}
\label{sec:DetAndShields}
\begin{figure} [!htb]
\centering
\includegraphics[width=0.6\textwidth]{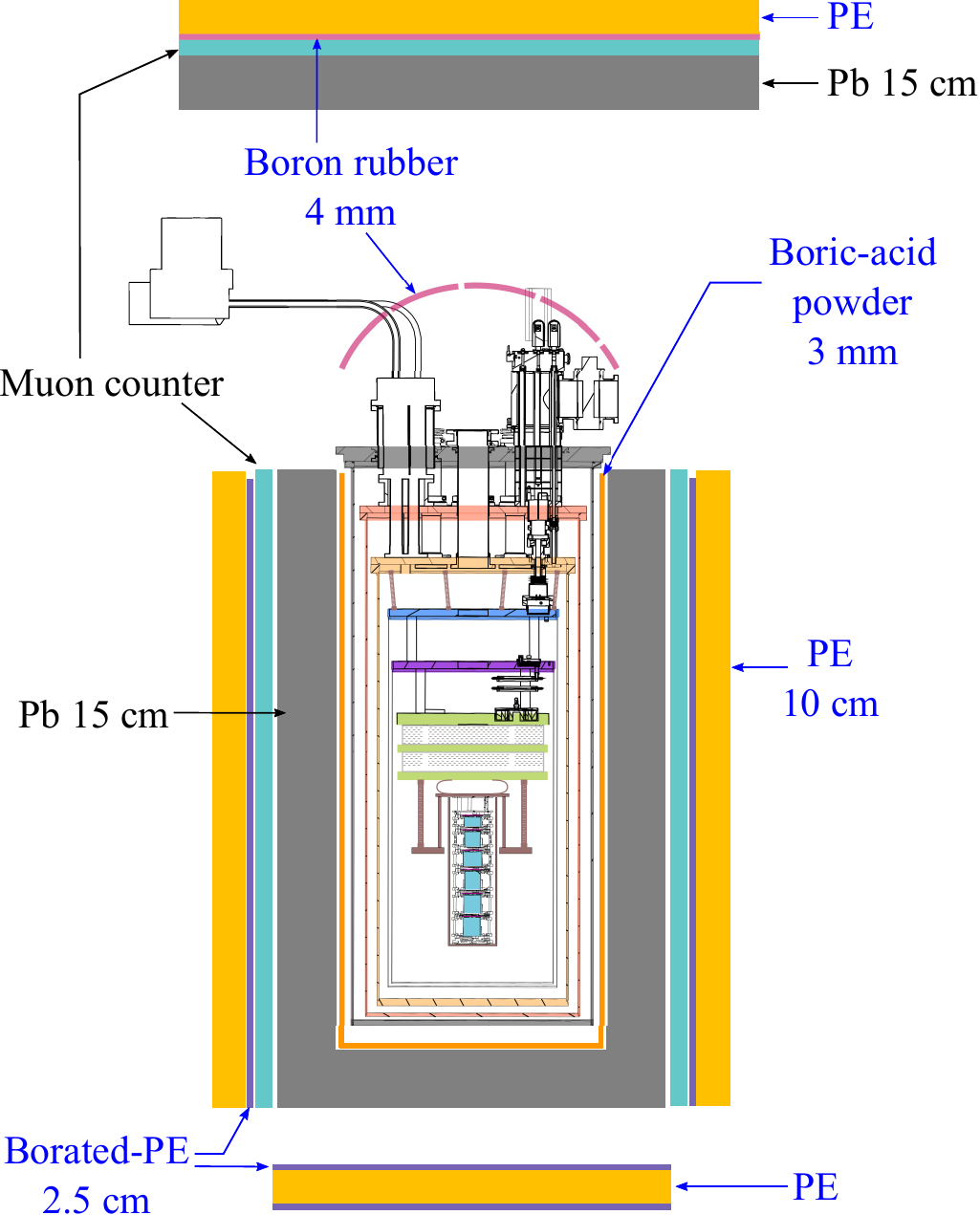}
\caption[]{Schematic view of the AMoRE-pilot detector system. Passive shields added for config-3 are denoted with blue arrows and texts.}
\label{fig:shields}
\end{figure}

In AMoRE-pilot, there were six CMO crystals, with calcium depleted in $\rm ^{48}Ca$ and molybdenum enriched in $\rm ^{100}Mo$, installed and used as the source of \znbb. $\rm ^{48depl}Ca$ was used to eliminate the potential background from $2\nu\beta\beta$ decay of \Ca. The CMO crystals have elliptical cylinder shapes, and their masses are 196~g, 256~g, 350~g, 352~g, 390~g, and 340~g for the detector modules from the top (CMO1) to bottom (CMO6), as shown in Fig.~\ref{fig:shields}. The average enrichment of $^{100}$Mo is 95\% with the differences between crystals being less than 1\%. The total mass of $^{100}$Mo in the six CMO crystals is 886~g. 

As a low-temperature thermal calorimetric detection with scintillating crystals \cite{Kim2022}, particle interactions with the CMO crystal induce an increase in temperature as well as scintillation photons. Both signals can be detected using metallic magnetic calorimeter (MMC) based detectors \cite{2015NIMPA.784..143Y}. 
Each AMoRE-pilot crystal detector module consisted of a CMO crystal held in a copper frame made of NOSV-grade copper from Aurubis AG~\cite{webNOSV}.
A phonon sensor was thermally well coupled on the elliptical bases of the crystal. On the opposite side, a photon sensor was installed.
A reflecting film (Vikuiti™ enhanced specular reflector from 3M™~\cite{webVikuiti}) was mounted in the copper frame to enhance the collection of the scintillation photon.
The crystal detector modules were assembled in a tower and attached to a mixing chamber plate of a dry dilution refrigerator with a thermal connection and a vibration mitigation system \cite{Kang:2017ezw,Lee:2017hya}. It was confirmed that the refrigeration system could reach well below 10 mK with a 1.6 $\mu$W cooling power under a long-term operating condition \cite{Kang:2017ezw}. A dedicated R\&D was brought forward to minimize the vibration of the detector modules resulting from the operation of the pulse tube refrigerator. Successful vibration suppression was obtained by a two-stage vibration mitigation system: a spring-loaded steel plate in the cryostat and a mass-spring damping system for the detector tower \cite{Lee:2017hya}.
The simultaneous measurements of the phonon and photon signals provided clear separation between \a~and \b/\g~events.
The principles and designs of the detector and sensors were described in \cite{,Kang:2017ezw,Alenkov:2015dic,Lee:2015tsa,Kim:2016xps,Kim:2017xrs}. 
Measurements with prototype detectors during commissioning operations showed an energy resolution of 8.7 keV FWHM at 2.6 MeV, and clear separations were demonstrated between \a~and \b/\g~events \cite{Kim:2017xrs}.

The detector tower was surrounded by a magnetic shield of a 2-mm-thick superconducting sheet made of low-radioactivity ancient lead. Furthermore, a 10-cm-thick ancient lead layer was placed between the detector tower and the mixing chamber plate as a radiation shield from the background due to the natural radioactivity of the materials composing the different temperature stages. 
Four layers of copper cans encased each temperature stage of the refrigerator system at temperatures of $\rm 50\ mK$, $\rm 1\ K$, $\rm 4\ K$, and $\rm 50\ K$, which were enclosed by a stainless steel container at room temperature.
During the measurement period in the final shielding configuration of the pilot phase, the detector system had a few layers of neutron shields installed with 10-cm-thick polyethylene (PE) bricks and 2.5-cm-thick borated PE plates, as shown in Fig. \ref{fig:shields}.

\subsection{Muon Veto System}
The muon veto counter system used in the AMoRE-pilot experiment consisted of ten plastic scintillator panels;
two panels ($76.2\times172.5~\textrm{cm}^{2}$) were placed on the top of the refrigerator, and eight panels ($55\times170~\textrm{cm}^{2}$~and \ $60\times170~\textrm{cm}^{2}$) were placed on the sides of the lead shielding. 
Cosmic muons generate sufficiently large signals in the 5-cm-thick plastic scintillator. 
Each scintillator panel was connected to 2-inch photomultiplier tubes (PMTs) via plastic light guides.
Depending on the spatial availability, two or three PMTs were mounted on one or two sides of the scintillator panels.
The muon counters covered 91\% if the solid angle around the crystal detectors.

\subsection{Data Acquisition}
Continuous signals from the detector modules' MMCs, both for the phonon and photon channels, were transferred to the analog-digital converter (ADC) called AMoREADC, sampled at every 10 $\mu$s with an 18-bit/channel resolution, and saved as raw data. Another ADC module (M64ADC) with a 62.5 MHz sampling rate was used for the muon counter system. A trigger-and-clock module board (TCB) provided a synchronized clock between AMoREADC and M64ADC modules. Data from the crystal detector modules and muon veto system were recorded separately but shared common timestamps.
Pulses in the raw AMoREADC data were selected by software using the Butterworth bandpass filter~\cite{Alenkov:2019jis,ctx6104737100003821}.

\subsection{Configurations of the Experimental Setup}
\label{Configs_of_the_Exp.setup}
During the AMoRE-pilot experiment, the detector system was improved twice to reduce the backgrounds. In this analysis, we compared the datasets from three configurations, referred to as config-1, config-2, and config-3.
During the AMoRE-pilot, calibration measurements were performed every week for a full day.

\begin{figure} [!b]
\centering
\includegraphics[width=0.9\textwidth]{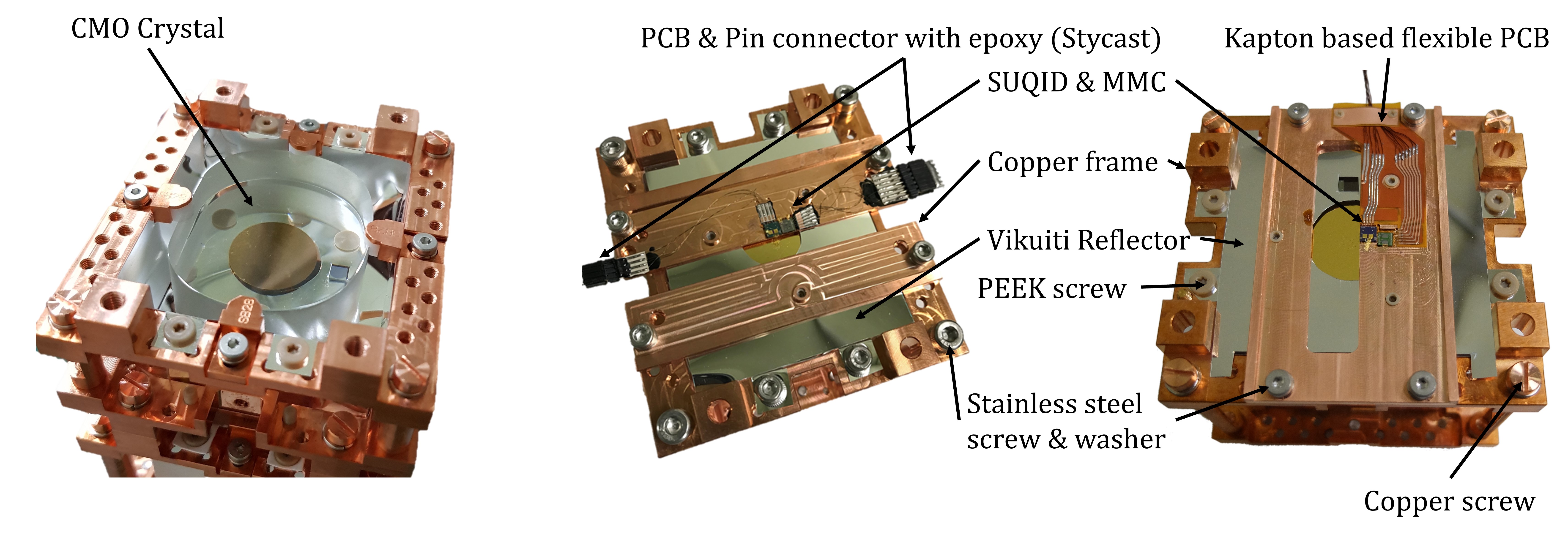}
\caption[Detectors] {A CMO crystal detector module before the light sensor assembly (left). Phonon sensor part of the detector module for in config-1 (left) and configs-2, 3(right). A more detailed description can be found in ~\cite{KIM2017105}.}
\label{fig:detector}
\end{figure}

\begin{figure}[!htb]
\centering
\includegraphics[width=0.99\textwidth]{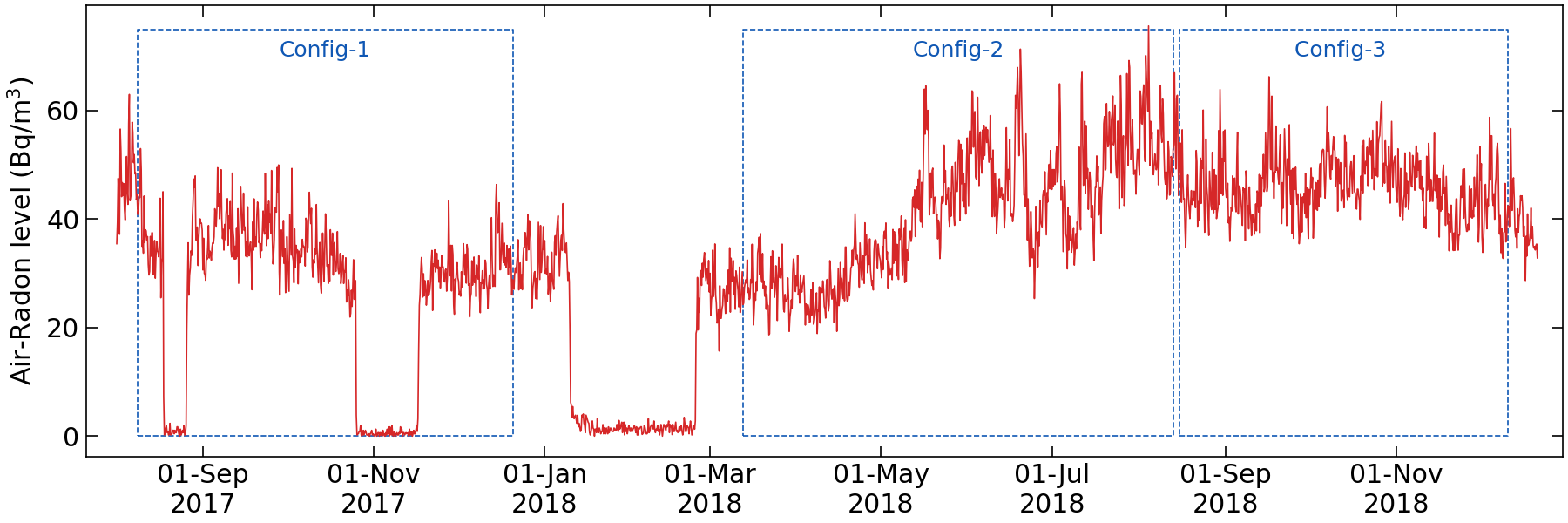}
\caption[Configurations and air-Radon level]{Data taking periods for different detector configurations and the radon level monitored in the neighboring laboratory room.}\label{fig:conf_rn}
\end{figure}

Figure \ref{fig:detector} shows the pictures of the detector modules used in config-1 and configs-2, and 3. 
In config-2, the wiring system of the detector module was redesigned using a Kapton-based flexible printed circuit board (PCB) \cite{hanwha} to replace the ceramic-based PCBs, the pin connectors, and the epoxy in config-1 that contain \U[238], \Th[232], and \K[40]~at levels ranging from hundreds of mBq/kg to a several Bq/kg.
Although their masses were small, they were placed near the crystals, and their contribution to the background was not negligible. 
More details regarding the background are discussed in Section~\ref{sec:results}. The soldering joints were made with pure lead-tin alloy in both configurations. Some of the stainless steel screws near crystals were replaced with custom-made polyether ether ketone (PEEK) screws or screws made of NOSV-grade copper.
Another important change in config-2, a stabilization heater was installed on two crystals to study the stability correction using heater pulse signals.
The wafer-holding springs of the photon detectors were made of phosphor bronze in config-1 but replaced with copper and polytetrafluoroethylene (PTFE) in config-2.
In config-3, the detector setup was kept as in config-2, but additional outer shielding of polyethylene (PE), borated-PE, and boric acid powder layers were installed to reduce the background induced by neutrons, as shown in Fig. \ref{fig:shields}.
The description of the configurations and the major change between them are summarized in Table~\ref{tab_runs}.
Responses of two detector modules went bad after removal of the pin connectors and reassembly for config-2, and two more detector modules became unusable after an accidental power outage during config-3.
Eventually, data useful for the analysis were acquired only for four and two detectors in configs-2 and 3, respectively.
Y2L is equipped with an air radon reduction system (RRS)~\cite{Ha:2022psk}, but it was not stable during the AMoRE-pilot, and radon-free air was supplied only for about 30\% of the config-1 data taking period as shown in Fig.~\ref{fig:conf_rn}
\begin{table*} [!htb]
\begin{center}
\caption{The run periods and corresponding detector configurations.}
\label{tab_runs}
\begin{tabular}{c|c|c|c|p{5cm}}
    \toprule
     Cfg. & Time  & Working CMOs & Temp. & Comments\\ \midrule
     1  & 135 days& all (1-6) & 20~mK & Crystal + muon veto system \\ 
     2  & 159 days & 2, 3, 4, 6 & 12~mK & Removal of near-crystal \newline radioactive components \\ 
     3  & 117 days & 2, 6 & 12~mK & Enhancement of neutron shielding \\ 
    \bottomrule
  \end{tabular}
\end{center}
\end{table*}

%% file: 3.dataAnalysis.tex
\section{Data Analysis}
\label{sec:data}

\begin{figure} [!ht]
\begin{center}
\includegraphics[width=0.9\textwidth, trim=0 0 0 0,clip ]{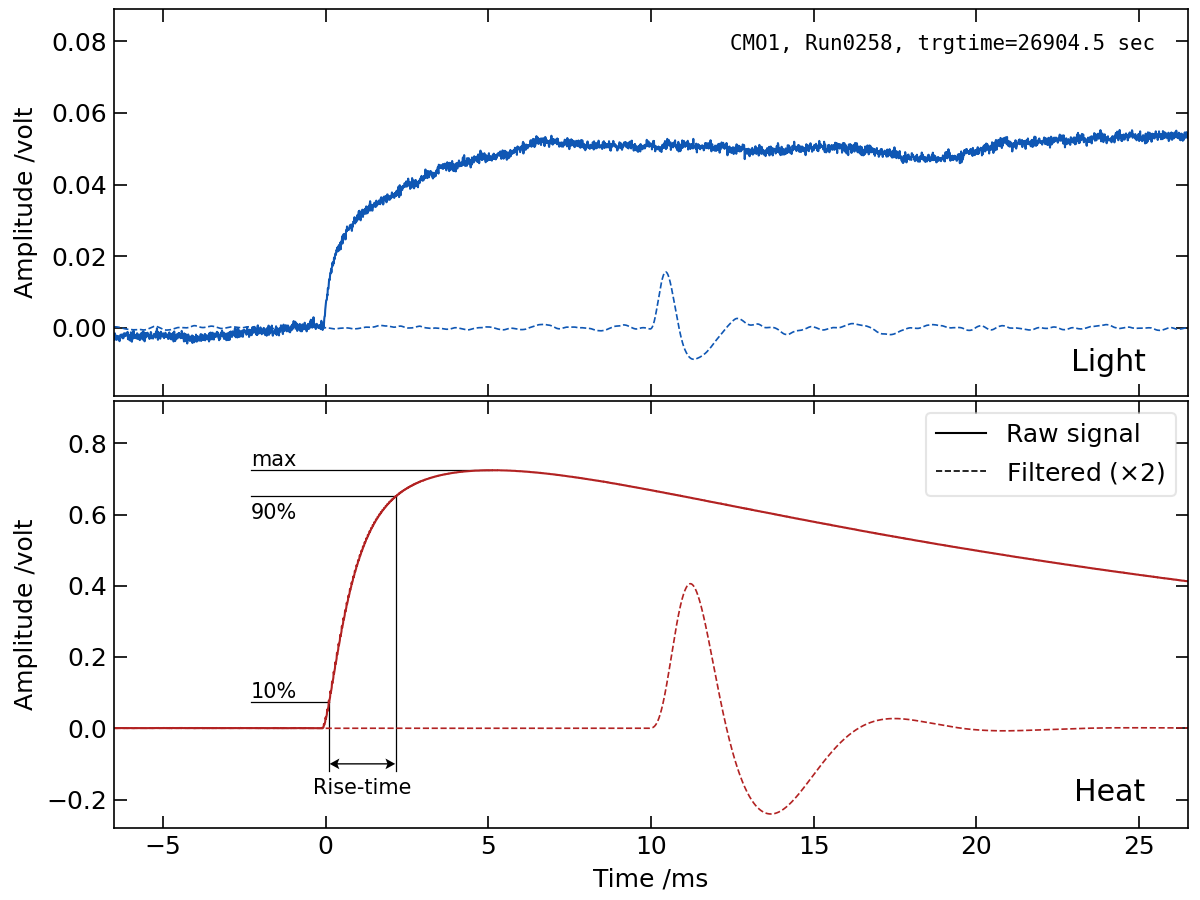}
\caption[wave_sample]{Example waveforms of the light (top) and the heat (bottom) signals of a 2.6-MeV $\beta/\gamma$ candidate event of CMO1. The solid lines denote the raw waveforms, and the dashed lines, which are drawn delayed by 10~ms, denote the waveforms after passing Butterworth bandpass filter.}
\label{fig:wave}
\end{center}
\end{figure}

\begin{figure} [!ht]
\begin{center}
\includegraphics[width=0.9\textwidth, trim=0 0 0 0,clip ]{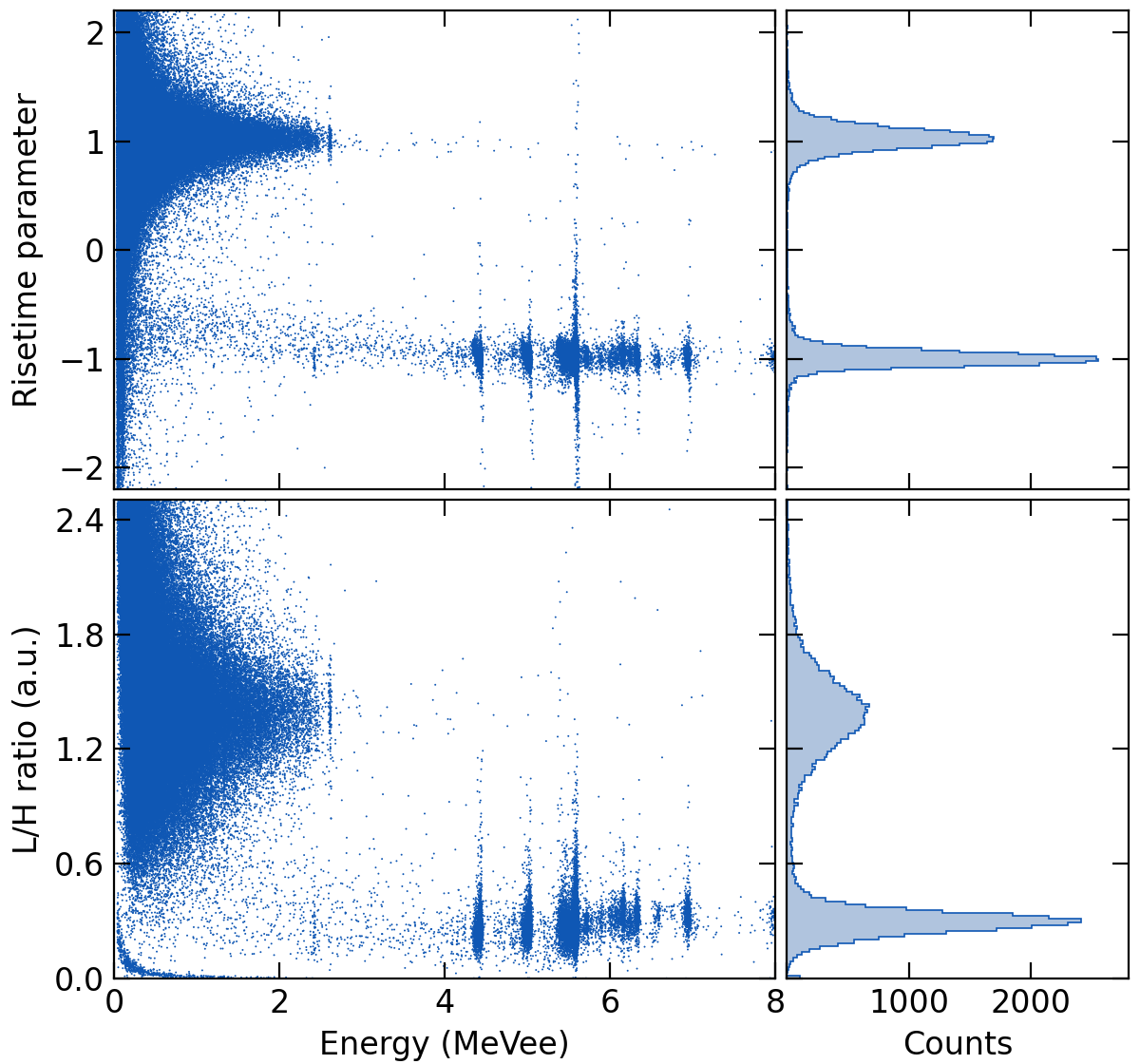}
\caption[PSD Parameters]{The rise-time and the $L/H$ parameters versus energy measured by the CMO1 detector in config-1. The rise-time and ratio of light to heat signals ($L/H$) were used to separate the \a~and \b/\g~events. Projections of the distributions for events with energy greater than 1 MeV are shown at right panels. Some events shown in the bottom-left corner of the $L/H$ versus energy plot are mainly due to the inefficient response of light detector for small signals at low energy.}
\label{fig:psd}
\end{center}
\end{figure}

Events were characterized by parameters calculated from their raw and bandpass-filtered waveforms, such as rise-time of the raw heat signals, amplitudes of the light ($L$) and heat ($H$) signals, and their ratio ($L/H$), as shown in Fig.~\ref{fig:wave}. The order and cutoff frequency of the Butterworth bandpass filter for each channel was selected for the best signal-to-noise ratio and energy resolution.
The rise-time is defined as the time difference between 10\% and 90\% of the raw waveform maximum.
The light signal amplitude is the difference between the maximum and the minimum of the bandpass-filtered waveform, while the heat signal amplitude was calculated by the least-square fitting of the filtered waveform to a template waveform.
The template waveform is an average of 2.6 MeV $\gamma$ event signals gathered from the calibration run.
The \a~and \b/\g~events were separated using the rise-time and the $L/H$, as shown in Fig. \ref{fig:psd}.
The rise-times were normalized for average values of  $\beta/\gamma$ and $\alpha$ events to be 1 and -1, respectively.
The $\beta/\gamma$ events were selected in the interval of two standard deviations from the means of rise-time and $L/H$ parameters.
Three anti-coincidence selections were applied to select the \znbb~decay event candidates: muon coincidence rejection, \a-tagging cuts, and single-hit selection. 
The first cut is to reject any events within 10 ms after the appearance of a muon candidate, defined when one or more muon counter panel have hit above the given threshold, to exclude events induced by muons. 
The \a-tagging cut was applied to any event within 15 minutes after a \Bi[212] \a-decay event candidate with an energy of 6207$\pm$50 keV to reject \b/\g~events from \Tl[208] decay to \Pb[208], which can produce background in the ROI.
The half-life of \Tl[208] is 3.05 minutes. A previous simulation study \cite{Luqman:2016okt} showed that a rejection window of 15 minutes removed 97.4\% of the background from \Tl[208] decays.
The single-hit selection cut discards events with hits in more than one crystal detector module.
More details about the data analysis are available in previous publications \cite{Alenkov:2019jis,Alenkov:2022}.

To build the energy spectra of selected \b/\g~ events, the mass of the crystal, the effective live time, and the total efficiency of each crystal detector module including muon rejection efficiency, \a-tagging efficiency, and single-hit efficiency were considered, as summarized in Table \ref{tab_effi}.
\begin{table} [!hb]
\begin{center}
\caption{Efficiencies and exposures in the configs-1, 2, and 3. \znbb~efficiency is estimated by MC simulation; analysis efficiency includes \b/\g~selection using rise-time and $L/H$, background uncertainty is inferred from the background model. Mass-time exposure includes \a-tagging and muon veto.}
\label{tab_effi}
\begin{tabular}{c|c|c|c}
    \hline
                              &  Config-1 & Config-2 & Config-3 \\ \hline
     \znbb~efficiency [\%]    &        \multicolumn{3}{c}{81.6} \\ \hline
     Analysis efficiency [\%] &      86.3 &     85.0 &     87.1 \\
     BKG uncertainty [\%]     &       5.4 &     10.3 &     21.2 \\
     Exposure [kg$\cdot$year] &      0.37 &     0.24 &     0.07 \\
    \hline
  \end{tabular}
\end{center}
\end{table}

%% file: 4.bkgSrcs.tex
\section{Background Sources}
\label{sec:bkgsrcs}

Background sources in the AMoRE-pilot were investigated using the energy spectra accumulated in the three configurations in an energy range from 1 to 8 MeV.
Potential background sources were identified as follows:
\begin{itemize}
    \item Decays of \K[40], \Th, \U[235], \U, and their daughter isotopes in the CMOs,
    \item \tnbb~decay of \Mo,
    \item Decays of \K[40], \Th, \U[235], \U, and their daughter isotopes in the detector set-up details and shielding,
    \item The \g~emissions from the surrounding rocks, and $\gamma$'s induced by muons and neutrons,
    \item The $\gamma$ emissions from the decay of $\rm ^{222}Rn$ in the air between the outer vacuum chamber (OVC) and Pb shield.  
    \item Radionuclides generated by cosmogenic activation in the detector materials, such as $^{60}$Co.
\end{itemize}

\subsection{Radioactive contamination of the CMO crystals and \tnbb~decay of \Mo[100]}
\label{sec:RadContOfTheCMO}
The activities of the \U, \U[235], and \Th~chains and their sub-chains in the CMO crystal scintillators were estimated by analyzing the \a~spectra measured in AMoRE-pilot \cite{Alenkov:2022}.
Sequences of decays starting with a long half-life and followed by decays of relatively short half-lives, shorter than a few days, are considered sub-chains. While long-lived isotopes can have varying activities due to chemical separation, decays within such sub-chains must be in equilibrium in most cases after detector construction. The \U, \U[235], and \Th~decay chains have four, three, and three sub-chains, respectively.
The activities of the sub-chains with \a~decays were estimated and are listed in Table \ref{tab_conc}.
\begin{table*} [!ht]
\begin{center}
\caption{Radioactive contamination of the CMO crystal scintillators [mBq/kg] \cite{Alenkov:2022}. The names of each crystal are given for identification, and the crystals masses are listed in Section \ref{sec:DetAndShields}.}.
\label{tab_conc}
{\footnotesize
\begin{tabular}{c|c|c|c|c|c|c|c}
    \hline
     Chain & Sub-chain & CMO1  & CMO2  & CMO3  & CMO4 & CMO5 &  CMO6 \\
    \hline \hline
     \multirow{ 5}{*}{\U} & \U  & 0.6(2) & 0.83(1) & 0.029(4) & 0.078(3) & 0.106(5) & 0.347(8) \\
     & \Th[230]  & 0.007(7) & 0.071(6) & 0.113(1) & 0.022(5) & 0.13(2) & 0.013(5) \\
     & \Ra  & 0.033(1) & 2.5(2) & 0.007(4) & 0.009(6) & 0.042(4) & 0.008(4) \\
     & \Pb[210]   & 3.9(7) & 183.1(2) & 0.16(1) & 0.77(2) & 3.81(5) & 0.61(2) \\\hline
     \multirow{ 3}{*}{\U[235]}& \U[235]  & 0.03(2) & 0.09(1) & 0.01(3) & 0.031(7) & 0.025(6) & 0.019(9) \\
     & \Pa[231]  & 0.003(5) & 0.021(5) & 0.003(1) & 0.003(1) & 0.012(2) & 0.002(2) \\
     & \Bi[211]  & 0.26(3) & 0.93(4) & 0.089(4) & 0.042(3) & 0.393(3) & 0.057(2) \\
     \hline
      \multirow{ 2}{*}{\Th}& \Th  & 0.005(3) & 0.033(3) & 0.001(1) & 0.001(2) & 0.001(1) & 0.004(2) \\
     & \Th[228]  & 0.006(2) & 0.156(2) & 0.0003(1) & 0.002(2) & 0.006(1) & 0.004(2) \\\hline
  \end{tabular}
}
\end{center}
\end{table*}

The \tnbb~expected rates in different crystals ranged from 155 to 308 events per day, estimated from the mass of \Mo~in the CMO crystals.
The \tnbb~decay of \Mo~is one of the dominant backgrounds up to $\rm 3\ MeV$. 
However, it did not contribute to the ROI for \znbb, except through random coincidence, which was insignificant \cite{Luqman:2016okt}.

\subsection{Material Radioassays}
\label{activity}
All the detector components and shielding materials were screened with either high-purity germanium (HPGe) detectors or inductively coupled plasma mass spectrometry (ICP-MS) \cite{2016JPhCS.718f2050S,GILEVA2023110673}. Several materials were measured using both methods.
The results of the HPGe measurements are listed in Table \ref{tab_act_hpge}, and those of the ICP-MS measurements are presented in Table \ref{tab_act_icpms}.

\begin{sidewaystable*} [!ht]
\centering
\begin{minipage}{\textwidth}
\centering
\caption{Activities of \U, \Th, and \K~in materials measured using HPGe~\cite{GILEVA2023110673}. The isotopes in parentheses represent the corresponding subchains. Components or materials that are major contributors in the $\beta/\gamma$ background modeling are denoted with asterisk (*). The unit is mBq/kg, if not specified.}
\label{tab_act_hpge}
\begin{tabular}{c|c|c|c|c|c|c}\toprule
  Components & \U & \Th & \Th & \multirow{2}{*}{\K} & Other isotopes & Config \\
  & (\Ra[222]) & (\Ac[228]) & (\Th[228]) & & \\\midrule\midrule
SQUID & \lt 0.01  & \lt 0.022 & \lt 0.01 &\lt 0.19 &  & All\\
Vikuiti Reflector & 0.6(2)  & \lt 0.9 & \lt 0.6 & 9(2) & & All\\
PEEK & \lt 1.80  & \lt 2.75 & \lt 1.18  & \lt 7.88 & & All\\
NOSV Cu \cite{Laubenstein:2009} & \lt 0.016 &  &\lt 0.025 & & & All \\ 
Stainless steel screw
& 2.1(4)  & \lt 3.9 & 4.7(6) & \lt 4.2 & \Co: 6.7(7) & All \\
Stainless steel plate
& 0.62(16)  & 0.84(29) & 0.82(18) & 2.2(14) & \Co: 2.1(3) & All \\ 
Rock$^{*}$ [Bq/kg] & 64.2(32) & 84.6(42) & 76.3(38) & 1581(79) & &  All\\ 
\midrule
Phosphor bronze holder & \lt 0.23  & \lt 0.40  & \lt 0.27  & \lt 1.31 & & 1 \\ 
Stycast 2850FT & 440(45) &  600(50) & 600(50) & 400(120) & & 1\\ 
PCB & 187(19) & 168(28) & 136(16) & 1060(223) & & 1\\ 
Pin connector$^{*}$ & 6110(170) & 7930(300) & 8070(290) & 8850(480) & & 1\\ 
\midrule
Kapton PCB  & \lt 1.1 & \lt 1.3 & \lt 1.1 & \lt 11.7  & & 2 \& 3 \\ 
PTFE  & \lt 0.5 & \lt 1.0 & \lt 0.8 & \lt 6.4 & & 2 \& 3 \\ 
Pb-Sn solder
& \lt 4 & \lt 13.5 &\lt 7.3 &  \lt 47.7  & \Pb[210]: 114(9) & 2 \& 3 \\ 
\midrule
Boric acid powder & \lt 1.83 & \lt 2.34 &\lt 0.69 & \lt 9.53  & & 3 \\
Borated rubber  & 626(34) & 748(46) & 2901(148) & 660(68) &  & 3 \\ 
\bottomrule
\end{tabular}
\end{minipage}
\end{sidewaystable*}

\begin{table} [!hb]
\begin{minipage}{\linewidth}
\centering
\caption{Concentrations of uranium (U) and thorium (Th) in the materials measured using ICP-MS~\cite{GILEVA2023110673}.}
\label{tab_act_icpms}
\begin{tabular}{c|c|c|c}\toprule
  & U & Th &  Config. \\ \midrule\midrule
Vikuiti Reflector (ppt) & 1.5(10) & 2.7(13) &  all\\
NOSV Cu\footnote{purchased in 2014} (ppt)& 1.66(4)  & 4.3(3) &  all \\ 
Rock (ppm) &  3.9(14)  & 11(7)   &   all \\ \midrule
Kapton PCB (ppt) & 893(90) & \lt 1 &  2, 3 \\ \midrule
\end{tabular}
\end{minipage}
\end{table}

\subsection{\g~ Emissions from the Surrounding Rocks and Neutrons}
Three rock samples, including shotcrete from cavity surfaces, were cut and measured by ICP-MS at the Korea Institute of Geoscience and Mineral Resource (KIGAM).
The results of activity measurements of the detector and shielding materials for the AMoRE experiment, including plastic materials such as PTFE and PEEK, soldering lead and glues will be published in a future publication under preparation.
Additionally, the environmental \g~spectrum without shielding was measured using an HPGe detector to validate the concentrations of \U, \Th, and \K[40] in the rocks.
The measurement results are shown in Table~\ref{tab_act_hpge}.

Radioactive decays from naturally occurring nuclides cannot account for most \b~and \g~events with energies exceeding 5 MeV. Possible background sources are muons passing through the crystals or secondary particles induced by muons. In addition, thermal neutrons captured by copper or iron in the shielding materials can generate high-energy \g's up to 8 MeV via ($n$,~\g) reaction.
The muon-related background events were negligible because of the high rejection efficiency by the muon anti-coincidence selection. 
The flux of thermal and fast neutrons in the Y2L was measured and employed for the {\sc geant4} simulation \cite{Yoon:2021tkv}.

\subsection{Cosmogenic Activation of Materials}
The cosmogenic activation by muons or neutrons is negligible in the experimental site, which has an overburden of about 2000-meter-water-equivalent. The activation of the experimental components by cosmic muons is primarily due to exposure of the experimental components at the above ground sea level during the preparation period. Especially, \Co~generated in the metallic materials has a relatively long half-life of 5.3 years and emits 1.173 and $\rm 1.333~MeV$ \g-rays which contribute to the background spectrum. Measured activities of \Co~in several metallic detector parts using HPGe detectors are shown in Table~\ref{tab_act_hpge}. 

%% file: 5.simulation.tex
\section{Simulations}
\label{sec:sim}

A background simulation was performed for the AMoRE-pilot detector, which includes the shielding configuration using the G{\sc eant4} simulation toolkit (version 9.6p02) \cite{Agostinelli:2002hh,Allison:2016lfl}.
The G4EmLivermorePhysics class was employed for low-energy electromagnetic processes such as the propagation of electrons, neutrons, muons, $\alpha$ particles, and heavy ions, including nuclear recoils from $\alpha$~emissions.
For the recoils from the \a\ emissions, the G4ScreenedNuclearRecoil was adopted for simulating their screening effects.
The generation of nuclear processes was based on the G4RadioactiveDecay database, built in the G\textsc{eant4} ~toolkit.
The DECAY0 program \cite{Ponkratenko:2000um} was used to generate \znbb~and \tnbb~decay events. A simulation software was developed to handle each radioactive decay as an event within a given measurement window in secular equilibrium.
Corresponding detector geometries have been developed for each configuration, and the simulation was performed separately.
The background event generated within a simulated component of the detector, whose physical dimensions are reflected, all relevant physics processed were simulated including potential deposition of any remaining energy of particles which may reach a crystal.
As in the real data analysis, the sequentially incident events within approximately $\rm 0.5\ ms$ for one crystal are considered to be indistinguishable from each other, so the two events are merged. A typical pulse width is approximately 30 ms. If consecutively incident events are between 0.5 and $\rm 100\ ms$, the second pulse is excluded. Additionally, the simulated events include multiplicity considering a $\rm 0.5\ ms$ time window for 6 crystals. 

Radioactive contamination in the crystal and \tnbb~decay are the dominant background sources \cite{Luqman:2016okt}.
Decays of \U, \U[235], \Th, and all daughters within their decay chains were simulated inside the CMO crystals, according to the composition activities determined in Section \ref{sec:RadContOfTheCMO}, they were normalized by sub-chains based on the measured activities from \a\ spectra in each crystal \cite{Alenkov:2022}.
The same selection cuts used for the background data processing were applied, including the \a-tagging cut and single hit selection.
The \tnbb~decay rate was simulated for the 95\% isotopic concentration of \Mo~in the CMO crystals.
Then, the simulated results were normalized to the expected rate in each crystal, taking into account the half-life of $^{100}$Mo $T^{2\nu}_{1/2}=7.12\,^{+0.21}_{-0.17}\times10^{18}$ yr \cite{Armengaud:2020}.

A realistic geometry for the detector module was implemented in the simulation.
Components located near crystals and any materials with relatively high contamination, such as Stycast™ epoxy glue, PCB, pin-connector, screws, and PEEK, were included.
When the experimental configuration was modified, the setup in the simulation was updated accordingly.

To study the contributions of radioactivity from environment to the background spectrum, the \K[40], \U,~and \Th~ full decay chains in the detector components and shielding materials were simulated.
Simulation of the whole process of rock $\gamma$'s--from generating them inside of rock surface surrounding the experimental hall to energy deposition at the target crystals, passing through all the materials in between-- at once consumed too much computing resources.
Instead, we first generated $\gamma$'s within 50 cm from the surface of the rock and retrieve the energy and angular distributions for the remaining $\gamma$'s at the surface (collection 1).
Then the $\gamma$'s sampled from the collection 1 and transferred to the inner surface of the lead shielding outside the OVC were selected.
The second collected spectra (collection 2) were built from these selected $\gamma$'s momentum and position distributions.
The energy deposit at the crystal was eventually simulated using collection 2. For the rock simulation, \a~ and \b~ particles were ignored since they could not reach the crystals.
The expected energy distributions at the crystal from the rock $\gamma$ were initialized to match the measured activities shown in Table~\ref{tab_act_hpge}. 

The \Co~background generated in the metallic materials by cosmogenic activation was simulated. However, the energy distributions from different detector components were almost indistinguishable. Therefore, the simulation result of the OVC, which is the most massive metallic component, was used to model the \Co~background effectively.

The background from ambient neutrons was simulated for different neutron shielding conditions, based on the measurement using dedicated neutron detectors~\cite{Yoon:2021tkv}.
The thermal and total neutron flux measured at the experimental hall were $(14.4 \pm 1.5)\times$$10^{-6}$~cm$^{-2}$s$^{-1}$ and $(44.6 \pm 6.6) \times$$10^{-6}$~cm$^{-2}$s$^{-1}$.
Additionally, we considered backgrounds from \Rn[222] in the air between the OVC and lead shielding. Specifically, some \g~emissions from decays \Bi[214], a descendant of \Rn[222], have energies over 3 MeV and can produce backgrounds at the \znbb~ROI.

%% file: 6-1.result.tex
\section{Results and Discussions}
\label{sec:results}

\subsection{Background Spectra}
\begin{figure*}[!h]
\centering
\includegraphics[width=1.0\textwidth]{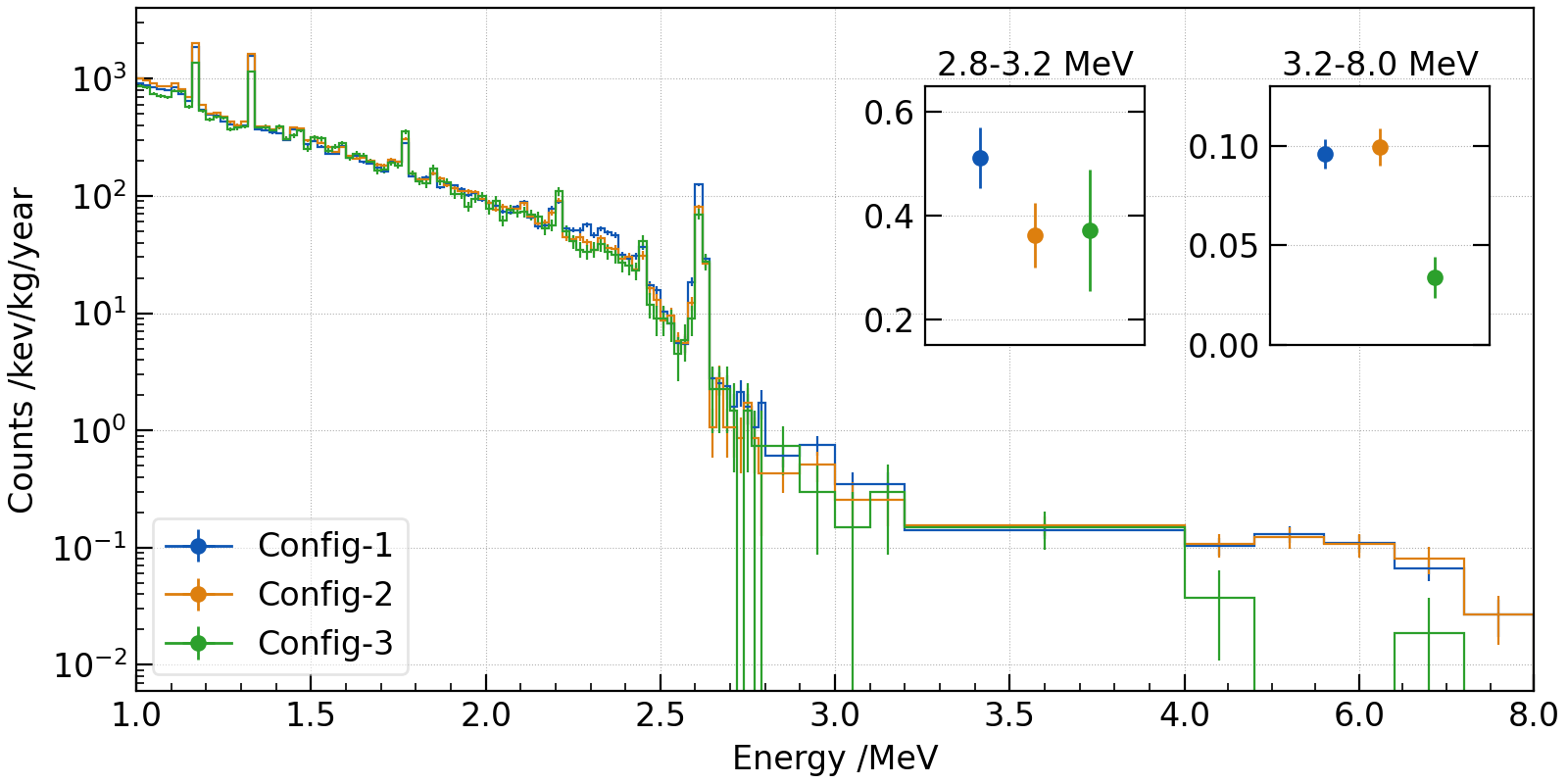}
\caption{$\beta/\gamma$ energy spectra of configs-1, 2, and 3, normalized over their mass-time exposures. Note that the visual bin sizes are shrunk above 4 MeV.
Changes in the background rate above 2.8 MeV are seen in the insets. In the two energy ranges, one near the ROI ($\rm 2.8-3.2\ MeV$) and the other mainly dominated by neutron induced background ($\rm 3.2-8\ MeV$), background rates decreased significantly between configs-1 and 2, and configs-2 and 3.}
\label{fig:sp567}
\end{figure*}

Figure \ref{fig:sp567} shows the measured $\beta/\gamma$ energy spectra in configs-1, 2, and 3, normalized by their mass-time exposures.
Several distinctive $\gamma$ peaks were identified as $\gamma$'s from the decays of the following radionuclides: \Tl[208]~(2615 keV), \Bi[214]~(1764 and 2204 keV), \Co[60]~(1173 and 1333 keV), and \K[40]~(1461 keV).
The 2.6~MeV $\gamma$ peak was shrinked by 40\% after config-1 by removing the near-crystal radioactive components such as pin-connectors and PCBs. The event rate in 2.8-3.2 MeV was also reduced by 27\% in configs-2 and 3.
In the energy range above 3.2 MeV, the config-3 spectrum showed a lower background level by 65\% compared to the configs-1 and 2 spectra, as expected by enhancement of external neutron shielding.

\subsection{Background Modeling}
The simulated $\beta/\gamma$ energy spectra, explained in section~\ref{sec:sim}, were fitted to the measured spectrum for each experimenal configuration, by floating the radioactivity of each background source.
In the fitting process, the initial activities were set based on the radioassay results shown in Table~\ref{tab_act_hpge}.
Not all the activities were freely floated, but some were fixed or constrained in the following manner:
\begin{itemize}
    \item Any component or material whose estimated contribution to the background rate was less than $10^{-5}$ counts/keV/kg/year in 1.0-2.8 MeV energy range was ignored.
    \item Among the near crystal components for config-1, the pin connectors were found to be the dominant contributor to the background and contribution of the others were far smaller. Because the shapes of the background spectra contributed from a common isotope origin were almost same among these near components, we fixed the minor components' activities and let only the pin connecter's activities float freely.
    \item Air-radon and rock-uranium also produced background shapes similar to each other and the both were major contributors to the background rate.
      Therefore, the air-radon activity for each configuration was softly constrained with the average and the standard deviation of the measured value shown in Figure~\ref{fig:conf_rn}.
    \item Ratios between activities of different decay chains from one source components were constrained to be consistent with the independent measurements in Table~\ref{tab_act_hpge}, within their uncertainty levels.
      It was to avoid the unnecessary correlation between activities of a same isotope in different source materials.
    \item Ratios of Rock and air-radon components' activities to the initial estimation were same for all experimental configurations.
    \item The internal radioactivity of the crystal was constrained by the $\alpha$ spectrum analysis result~\cite{Alenkov:2022}, and the \tnbb~decay half-life was constrained at 7.1$\pm0.2\times10^{18}$ years.
\end{itemize}
The reason for these restrictions is to prevent failure in fitting mainly caused by unnecessary correlations among activities of different source components that gave similar background spectrum shapes.

\begin{figure} [!htb]
\centering
\includegraphics[width=0.95\textwidth]{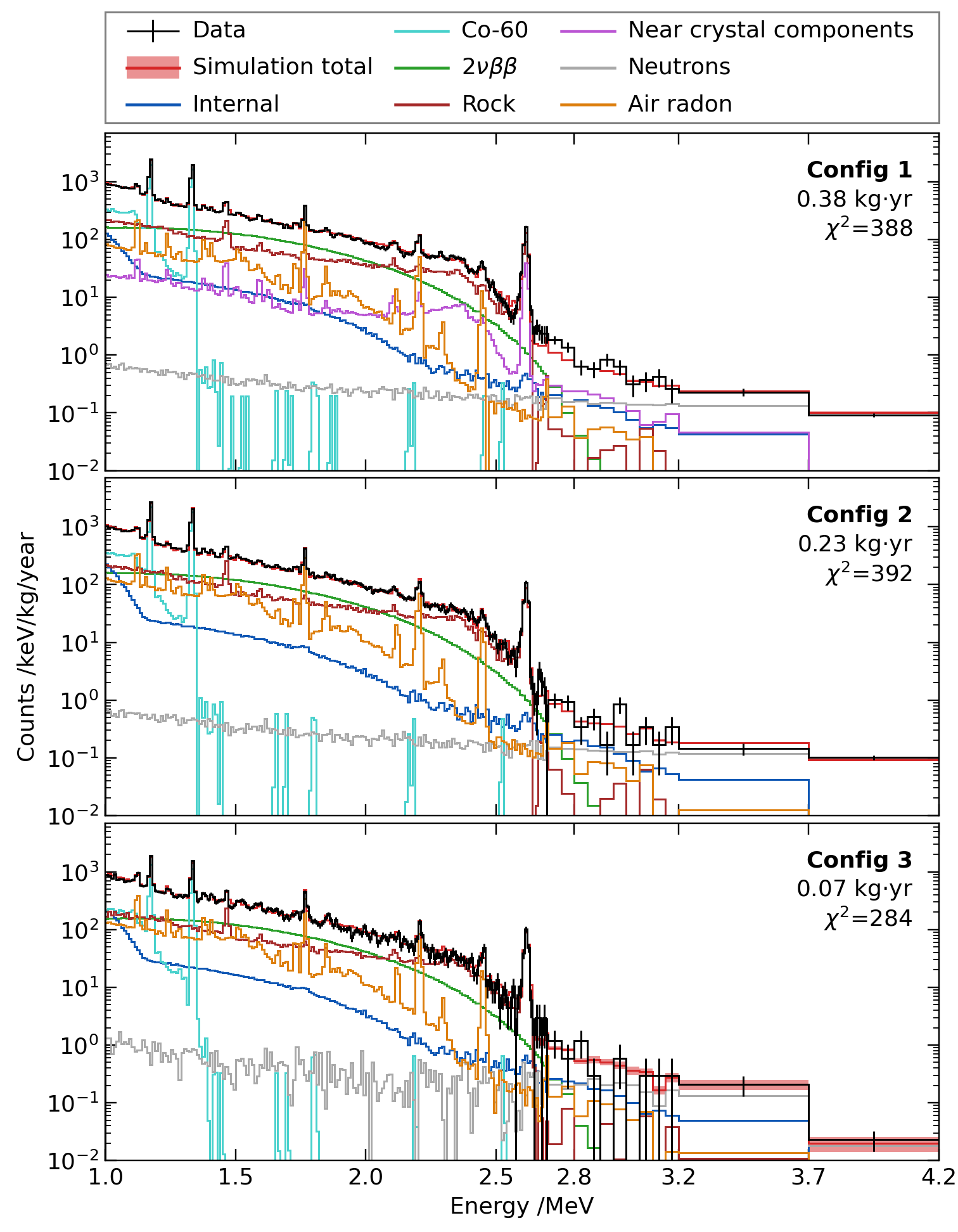}
\caption[Background modeling result]{
Energy spectra of $\beta$/$\gamma$ events measured in three different configurations of the AMoRE-pilot setup. The experimental data are represented by black histograms with errorbars, the fitting result by red histograms with errorbands, and the components of the background model by other colors. The total number degree of freedom is: [numbers of bins with non-zero data]-[number of free parameters]=544-9=535.
}
\label{fig:run5MCfitto4}
\end{figure}

The resulting background models overlaid with the measured spectra for all three configurations are shown in Fig.~\ref{fig:run5MCfitto4}, and the fitted radioactivities and their ratios to the initial estimations are shown in Table~\ref{tab_fit_conf}.
Activities of all decay chains for the pin connector in Config-1 fit about half of the initial estimation, but the ratios between the chains were more consistent than the given constraints given by uncertainty levels of HPGe measurements.
We interpreted this was mainly due to imperfact location of this pin connector component in the simulation which could cause a large uncertainty.
The Air-radon component were fitted to be about 50\% higher than the independent measurement in the neighboring room.
This was acceptable since the air flow could be different between the laboratory rooms.
The \Co~component was found to become smaller in config-3, indicating there could be other significant source components than OVC but they could not be identified in this study.
\begin{table*} [!hb]
\caption{Fitted radioactivities of the materials in configs-1, 2, and 3. Ratio of the post-fit result to the initial estimation is shown in the square bracket. Activities of rock and air-radon components are fully correlated between configurations. The activities are expressed in units of $\rm Bq/kg$, except for air-radon in Bq/m$^{3}$.}
\label{tab_fit_conf}
\centering
{\footnotesize
\begin{tabular}{c|c|c|c|c}\toprule
\multirow{2}{*}{Component}
& Decay & Config-1 & Config-2 & Config-3 \\\cmidrule{3-5}
         & chain & Activity [Ratio]    & Activity [Ratio]   & Activity [Ratio]\\\midrule\midrule
\multirow{3}{*}{Rock} 
& $^{222}$Ra  & \multicolumn{3}{c}{63.1(31) [0.98(5)]} \\
& $^{228}$Th & \multicolumn{3}{c}{69.0(13) [0.90(2)]} \\
& $^{40}$K   & \multicolumn{3}{c}{1672(52) [1.06(3)]} \\ \midrule
& $^{222}$Ra & 3.11(22) [0.51(4)] & - & - \\
Pin& $^{228}$Ac & 4.03(35) [0.51(4)] & - & - \\
connectors & $^{228}$Th & 4.24(29) [0.52(4)] & - & - \\
& $^{40}$K   & 4.64(43) [0.52(5)]  & - & - \\ \midrule
Air & $^{222}$Rn & 38.7(8) [1.47(3)] & 61.0(12) [1.47(3)] & 66.8(13) [1.47(3)]\\ \midrule
OVC & $^{60}$Co & 0.110(1) & 0.118(1) & 0.070(2) \\
\bottomrule
\end{tabular}
}
\end{table*}

In the energy range 1.0-2.8 MeV, dominant sources of background were \tnbb, rock-$\gamma$'s, and air-radon for all configurations. Particularly in config-1, near crystal components represented by the pin-connectors also possessed a large portion of background.
After removing these near components for configs-2 and 3, the background rate from them were reduced by two orders of magnitudes.
Reduction of the 2.6 MeV $\gamma$ peak in configs-2 and 3 was also explained by the removal of the near components.
Major contributions to the background rates for energy range around Q-value, from 2.8 to 3.2 MeV, in different experimental configurations are shown in Table~\ref{tab_bkg_roi}.
It was confirmed that the reduction of background rate was also due to removal of the near components after Config-1.
Next to that, the internal and surface radioactive contaminant of the crystals dominated, having contributions as large as 0.2 counts/keV/kg/year.
Especially, CMO 2, which showed the highest internal background rate among the AMoRE-pilot crystals~\cite{Alenkov:2022}, had the most significant background in this energy range.
Neutron-induced background was identified to be another large contributor here, and to be the most significant one for the higher energy range.
Enhancement of neutron shielding for config-3 was found to be significantly effective for the high energy range above 3.7 MeV, but below that almost no reduction was found.
The other remaining background around ROI was by $\gamma$'s from the decays of $^{214}$Bi ($Q_{\beta}$=3272 keV) in the air and rock, and $^{208}$Tl ($Q_{\beta}$=4999 keV) in the rock.
\begin{table} [!hb]
\centering
\caption{Major background contributions in the energy range of 2.8-3.2 MeV, in units of counts/keV/kg/year.}
\label{tab_bkg_roi}
\begin{tabular}{c|c|c|c}\toprule
  & Config-1 & Config-2 &  Config-3 \\ \midrule\midrule
Internal & 0.097(2) & 0.113(2) & 0.124(3) \\
Rock & 0.018(1) & 0.017(1) & 0.020(2) \\
Near crystal component & 0.139(23) & - & - \\ 
Neutrons &  0.146(2)  & 0.129(2) & 0.189(190) \\
Rn in air & 0.034(1) & 0.052(2) & 0.060(2) \\ \midrule
Subtotal & 0.434(11) & 0.311(8) & 0.393(62)\\\midrule
Data & 0.512(58) & 0.362(62) & 0.372(117) \\\bottomrule
\end{tabular}
\end{table}

%% file: 6-2.discussion.tex
\subsection{\znbb~half-life}
\begin{figure} [!htb]
\centering
\includegraphics[width=1.0\textwidth]{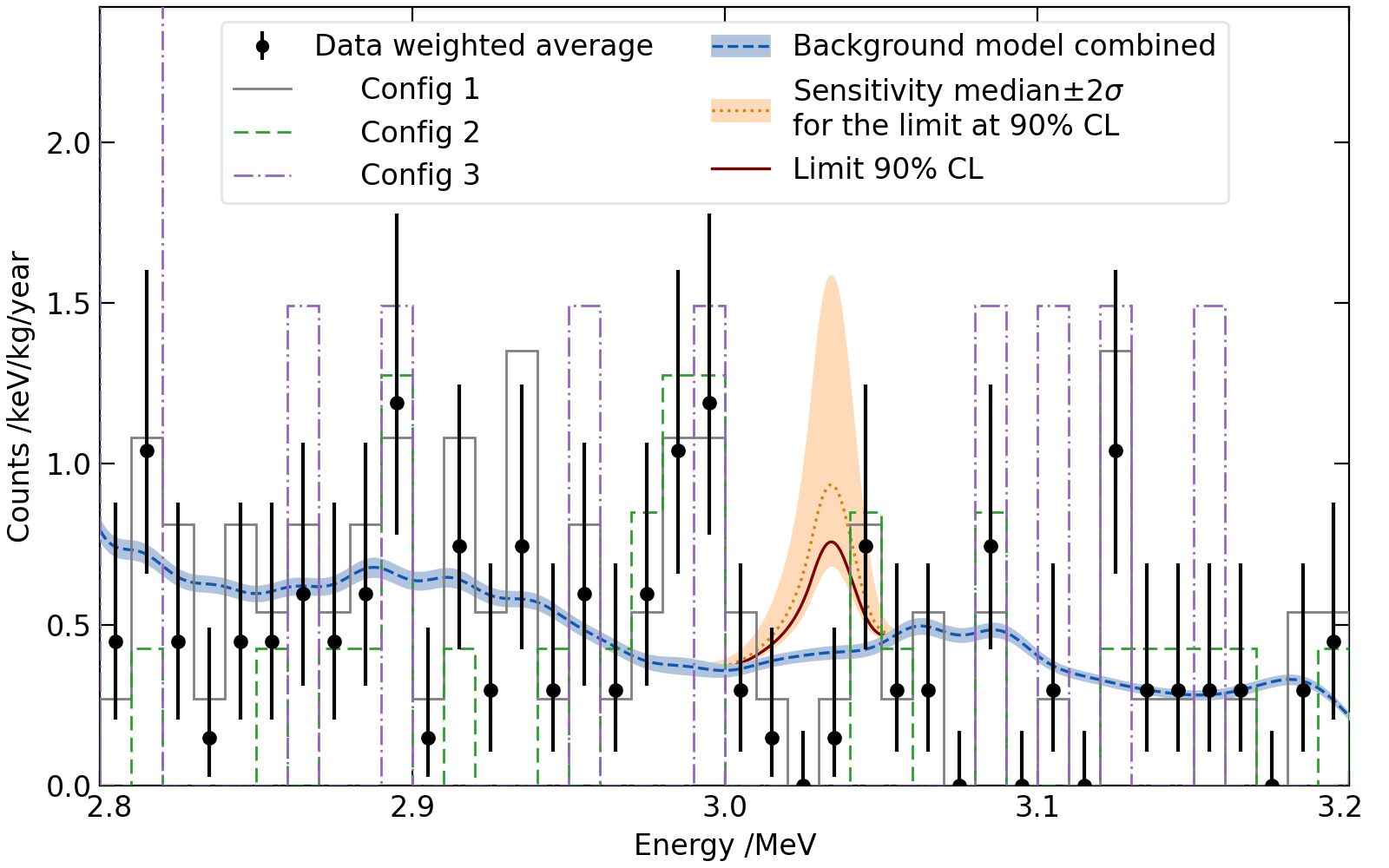}
\caption{The AMoRE-pilot energy spectrum and the model background around the region of interest ($E=$2.8-3.2 MeV). The combined data for all 3 configurations with the poisson errors are shown as the black dots and error bars. Data from configurations-1,2, and 3 are denoted by the solid-gray, dashed-green, and dash-dotted-purple histograms, respectively. The combined model background and its uncertainty are depicted as the blue dashed-curve and band. The \znbb~decay signal expected with the half-life at the upper limit of 90\% confidence level (CL) is shown as the dark-red curve. The median and 2-$\sigma$ range of the sensitivity for the upper limit of the \znbb~half-life given the model background are shown as the orange dotted curve and band.}
\label{fig:0vbb_result}
\end{figure}

In this study, the total data for 0.68 kg$\cdot$year exposure of AMoRE-pilot was analyzed, including the 0.3 kg$\cdot$year exposure data of our previous report \cite{Alenkov:2019jis}. With the larger amount of data and better understanding of the backgrounds, a new limit on the \znbb~decay of \Mo~was calculated. 
The signal shape was modeled in analogous to the peak at 2615 keV ($^{208}$Tl), and the resolution was extrapolated using distinctive $\gamma$ peaks in the background spectrum. The model background spectrum around ROI was converted to a smooth and continuous function for an unbinned analysis using the kernel density estimation~\cite{10.1214/aoms/1177704472}. The model function comprised four parameters: the signal and background sizes, and the signal energy and its resolution obtained from energy calibration. An unbinned negative log likelihood was minimized and profiled for the non-negative signal size. The likelihood function $\mathcal{L}$ was defined as:
\begin{equation}
\label{eq:likelihood}
    \mathcal{L} = \frac{n_e^{n_o} \cdot e^{-n_e}}{n_o !} \cdot \pi_b \cdot \prod_{i=1}^{n_0} p(E_i),
\end{equation}
where $n_e$ and $n_o$ are the numbers of expected and observed events in the ROI, respectively, $\pi_b$ is the Gaussian background constraint, and $p(E_i)$ is the probability density of the model function at energy $E_i$.
To take into account the systematic uncertainties, we put Gaussian constraints on the size of the background, \znbb~peak location, and energy resolution. The uncertainties on the background sizes for configs-1, 2, and 3 were 3.4\%, 9.6\%, and 25.0\%, respectively, assigned from the background modeling result. The uncertainties on the peak location and energy resolution were estimated from the $\gamma$ peaks in the background spectrum.
Each configuration (1, 2, and 3) was fitted separately. All results were combined considering efficiencies and mass-time exposures, as summarized in Table \ref{tab_effi} and Fig. \ref{fig:0vbb_result}. A new \znbb~half-life limit of \Mo~was obtained as $3.0 \times 10^{23}$ years in 90\% C.L. The 2$\sigma$ band around the sensitivity median, assuming no \znbb, is obtained from the pseudoexperiment as $[0.7, 3.5] \times 10^{23}$ years. 
The \znbb~half-life of \Mo~presented recently at CUPID-Mo was $1.8 \times 10^{24}$ years at 90\% C.L., as obtained from a measurement of 4.2 kg of enriched $\rm Li_{2}$$\rm^{100}MoO_{4}$ (LMO) crystals for $^{100}$Mo exposure of 1.47 kg$\cdot$year \cite{Augier2022}.

The next stage, AMoRE-I data taking, is finished with \sm $\rm 4.6\ kg$ of CMO and \sm $\rm 1.6\ kg$ of LMO crystals in the same cryostat as in the pilot stage at Y2L. The \Rn[222] background in the room where the detector system is installed has been reduced about two orders of magnitude lower than the pilot stage (\sm 0.05 $\rm Bq/m^3$ to \lt tens of $\rm \mu Bq/m^3$). In addition, the minimum thickness of the neutron shielding was increased from 10-30 cm to 20-30 cm, and the lead shielding was increased from 15 cm to 20 cm.
The additional 5-cm-thick lead shielding reduces \g's in the range of 2 to 3 MeV by orders of magnitude. Furthermore, to minimize the untagged muon events due to incomplete coverage, the muon veto system was upgraded to cover the sensitive detector volume fully.
Furthermore, detector design modifications have been made to reduce materials near the crystals.
The final stage, AMoRE-II, will feature $\sim$100 kg of \Mo~ isotope in the form of LMO crystals at a new underground laboratory \cite{Park_2021}. The AMoRE-II experiment is expected to reach a half life sensitivity of $10^{26}$ years with $\sim$100 kg of $\rm ^{100}Mo$ in five years.

%% file: 7.summary.tex
\section{Summary}
\label{sec:summary}

The AMoRE experiment aims to search for \znbb~ using \Mo. In the pilot experiment, we used six CMO crystals enriched in $^{100}$Mo and depleted in $^{48}$Ca with a total mass of 1.9 kg in a series of runs. We upgraded the detector system as data-taking proceeded. The measured spectra were modeled using extensive MC simulations, and the main background sources were identified. We reduced the background by removing contaminated materials near the crystals and installing neutron shields with boric acid powder. 

The background between $\rm 2.8\ MeV$ and $\rm 3.2\ MeV$ in the last configuration of the pilot stage was $\rm 0.38\ ckky$.
The next stage, AMoRE-I experiment, has completed data taking with LMO crystals installed along with the CMO crystals. The detector system upgrades developed in AMoRE-pilot have resulted in an order of magnitude reduction in background rates in the ROI for AMoRE-I.
The AMoRE-I aims to reduce the background further and estimate the requirements for the AMoRE-II stage to achieve a background level of less than $\rm 10^{-4}\ ckky$ needed to realize a zero-background experiment able to probe the inverted hierarchy of the neutrino mass pattern.